\documentclass[aps,prb,superscriptaddress]{revtex4-2}
\usepackage{graphicx}
\usepackage{bm}
\usepackage{natbib}
\usepackage{color}
\usepackage{epstopdf}
\usepackage{amsmath}
\usepackage{url}
\usepackage{amssymb}
\usepackage{pifont}
\usepackage{gensymb}
\usepackage{braket}
\usepackage{blkarray}
\usepackage[utf8]{inputenc}

\begin{document}


\title{Topological phases of an interacting Majorana Benalcazar-Bernevig-Hughes model}

\author{Alfonso Maiellaro}
\affiliation{Dipartimento di Ingegneria Industriale, Universit\`{a} degli Studi di Salerno, Via Giovanni Paolo II, 132, I-84084 Fisciano (SA), Italy}
\author{Fabrizio Illuminati}
\affiliation{Dipartimento di Ingegneria Industriale, Universit\`{a} degli Studi di Salerno, Via Giovanni Paolo II, 132, I-84084 Fisciano (SA), Italy}
\affiliation{INFN, Sezione di Napoli, Gruppo collegato di Salerno,Italy}
\author{Roberta Citro}
\affiliation{INFN, Sezione di Napoli, Gruppo collegato di Salerno,Italy}
\affiliation{Dipartimento di Fisica "E.R. Caianiello", Universit\`{a} degli Studi di Salerno, Via Giovanni Paolo II, 132, I-84084 Fisciano (SA), Italy}
\affiliation{CNR-SPIN, Via Giovanni Paolo II, 132, I-84084 Fisciano (SA), Italy}
\date{\today}
\begin{abstract}
We study the effects of Coulomb repulsive interactions on a Majorana Benalcazar-Bernevig-Huges (MBBH) model. The MBBH model belongs to the class of second--order topological superconductors ($HOTSC_2$), featuring robust Majorana corner modes. We consider an interacting strip of four chains of length $L$ and perform a Density Matrix Renormalization Group (DMRG) numerical simulation based on a tensor--network approach. Study of the non--local fermionic correlations and of the degenerate entanglement spectrum indicates that the topological phases are robust in the presence of interactions, even in the strongly interacting regime.
\end{abstract}
\pacs{Valid PACS appear here}
\maketitle

\section{Introduction}
\label{2DSSHMajoranas}
In recent years, higher--order topological superconductors (HOTSCs) have attracted significant interest and have been investigated in depth as novel platforms to realize topological superconductivity \cite{PhysRevLett.111.047006,PhysRevLett.118.147003,PhysRevB.97.205136,PhysRevLett.121.196801,PhysRevB.100.054513}. Aside from their theoretical interest, such systems are potentially relevant for applications; in particular, devices based on two--dimensional HOTSCs hosting Majorana corner states have been proposed to implement braiding dynamics \cite{PhysRevResearch.2.032068} and quantum gates for quantum computation \cite{PhysRevA.52.3457,PhysRevB.88.035121,Lian10938,PhysRevA.62.052309}. HOTSCs have surface states that propagate along one-dimensional lines (hinges) or are localized at some points (corners) on the surface. In particular, for such systems $m$-dimensional Majorana corner states can be realized in $d$-dimensional superconductors, with $m \leq d-2$.\\

In Ref.~\cite{condmat6020015}, two of us have introduced a model of second-order topological superconductor ($HOTSC_2$) based on Majorana fermions (Mfs) operators. The model is the equivalent of the Benalcazar--Bernevig--Hughes (BBH) model \cite{doi:10.1126/science.aah6442} for Dirac fermions. In the Majorana BBH model (MBBH), $C4$ symmetry and reflection symmetries ensure robustness of the corner states. In fact, the model belongs to the trivial two-dimensional $BDI$ class satisfying time-reversal, particle-hole and chiral symmetries \cite{PhysRevB.55.1142}. Moreover, the crystalline symmetries ensure a quantized two-dimensional Zak phase \cite{https://doi.org/10.1002/pssb.202000090,RevModPhys.66.899,PhysRevLett.118.076803}. When written in terms of Dirac fermions, the MBBH model is also equivalent to a model of Kitaev chains coupled by a staggered pairing coupling.\\

On the other hand, one of the main challenges in the study of quantum matter concerns the robustness of topological phases and topological superconductivity in the presence of interactions. The question has been addressed by considering the effect of Coulomb repulsive interactions \cite{PhysRevX.7.031057,PhysRevB.84.014503,PhysRevX.2.031008} and/or by developing a number--conserving theory \cite{PhysRevLett.111.173004} in order to go beyond the BCS approximation. In particular, some recent studies have investigated the physical properties of HOTSCs when Coulomb interactions are taken into account \cite{10.1088/2053-1583/ac4060,scammell2021intrinsic}.\\
The mean-field approximation offers a "cartoon" description of the topological phases within the single particle approximation. On the other hand, adding interactions removes the single-particle approximation and opens the way to the investigation of more realistic settings. Thus motivated, in the present work, by means of a density matrix renormalization group (DMRG) numerical analysis, we consider the effects of repulsive Coulomb interactions on a specific version of the MBBH model, namely an interacting strip with four chains. The strip setting is particularly interesting because it realizes the minimal model able to characterize the crossover from a $1D$ to a $2D$ system. It can thus be reduced to an effective one--dimensional lattice, that is a suitable lattice for simulations with DMRG technique implemented by matrix product states (MPS) and, on the other hand, it can shed light on the topological robustness expected when one considers fully two--dimensional geometries.\\
The results of our analysis careful analysis suggest that the quasi one--dimensional system supports two robust zero energy fermionic modes localized at the two edges of the strip and robust even in the limit of strong interactions. Although the topology of the system is intrinsically two--dimensional, the quasi one--dimensional case provides signatures of the robustness of the topological regime against repulsive interactions. In particular, these robust fermionic modes are expected to split into corner Majorana modes for a fully two--dimensional square-lattice system with length $L$ equal to the width $N$. The twofold degenerate entanglement spectrum, reported in Section \ref{DMRGInteracting}, is a further possible signature of topological order. A complete proof of such a robustness requires to test the system in the fully 2D geometry using three tensor networks, and we expect to accomplish this task in the near future.\\

Despite being characterized by a specific set of parameters, our system could be realized within the framework of network models \cite{PhysRevB.103.115428,PhysRevB.89.075113,PhysRevLett.125.096402,PhysRevB.99.045441} instead of traditional condensed matter setups. The experimental implementation of the former has been successfully realized using meta-materials platforms, as optical fibers and coupled ring resonators \cite{PhysRevB.100.085138,RevModPhys.91.015006,PhysRevLett.110.203904}. Theoretical proposal of network models have already been introduced to describe some condensed matter systems made of chiral Majorana modes \cite{PhysRevB.103.115428} or to describe physical phenomena such as the quantum Hall effect \cite{Chalker_1988}. Network models enable a large degree of control on the parameters subject to various constraints, including particle-hole symmetry. In fact, network models with particle-hole symmetric spectra have already been probed in recent experiments~\cite{PhysRevLett.124.253601,Gao}.\\

The paper is organized as follows. In Section \ref{2DModel}, we review the main properties of the mean-field MBBH model introduced in Ref.~\cite{condmat6020015}. In Section \ref{InteractingStrip}, we study the particular case of an interacting four-chain strip; in subsection \ref{AnalyticalResults}, we discuss the analytical operations needed to transform the strip into an effective one-dimensional system and to perform the DMRG simulations with MPSs. In subsection \ref{DMRGInteracting} we discuss the main numerical results. We track the fermionic correlations along the strip geometry and we study the trend of the entanglement spectrum from the perturbative to the strongly interacting regime. Conclusions are drawn in Section \ref{Conclusions}. Appendix \ref{AppendixA} contains details on the matrix product operator (MPO) tensors built for our model.
\section{Majorana BBH model }
\label{2DModel}
In this section, we briefly discuss the main topological properties of the BBH model. This provides the appropriate starting point before treating the case of the interacting four-chain strip. Following Ref.~\cite{condmat6020015}, we consider a system of Majorana fermions with staggered couplings (w, v) confined in a two--dimensional lattice and described by the Hamiltonian:
\begin{eqnarray}
	\begin{aligned}
		H_0=&\frac{i}{2}\biggl[w\sum_{m,l=1}^{L,N}a_{m,l}b_{m,l}+v\sum_{m,l=1}^{L-1,N}b_{m,l}a_{m+1,l}+w \sum_{l=1}^{N-1,2} \sum_{m=1}^{L} \biggl( b_{m,l}b_{m,l+1}-a_{m,l}a_{m,l+1}\biggr)\\
		&+v\sum_{l=2}^{N-1,2} \sum_{m=1}^{L} \biggl( b_{m,l}b_{m,l+1}-a_{m,l}a_{m,l+1}\biggr)\biggr].
	\end{aligned}
	\label{MajoranaH}
\end{eqnarray}
$a_{m,l}$ and $b_{m,l}$ are the Majorana operators belonging to a complex fermion operator $c_{m,l}=(a_{m,l}+ib_{m,l})/2$. $L$ and $N$ are, respectively, the length and the width of the system, $m$ and $l$ are the lattice sites. The synthetic $\pi$ flux per plaquette gives rise to the staggered couplings via the Peierls substitution \cite{PhysRevLett.118.076803}.
\begin{figure*}
	\centering
	\includegraphics[scale=0.30]{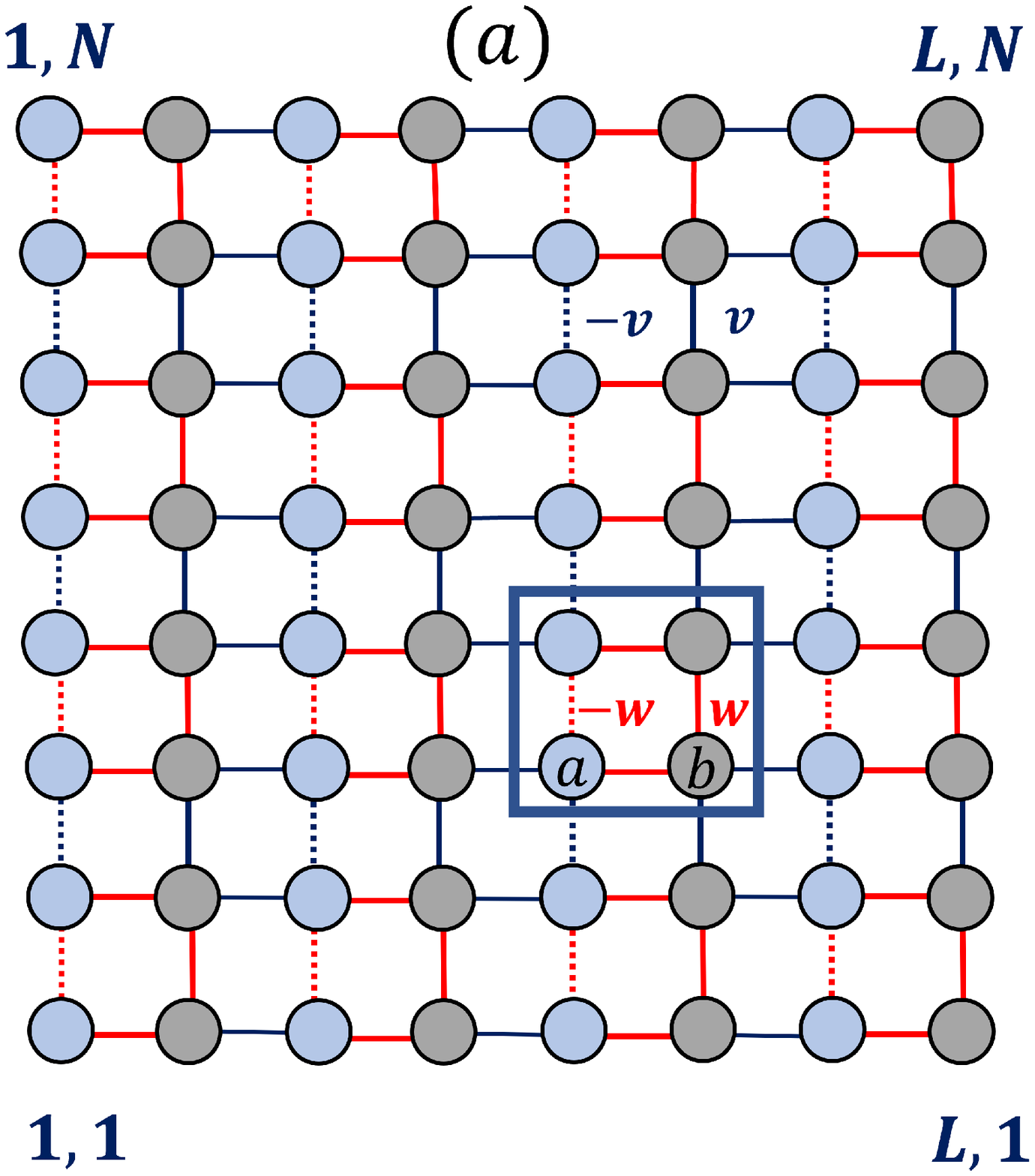}
	\hspace{0.5cm}
	\includegraphics[scale=0.32]{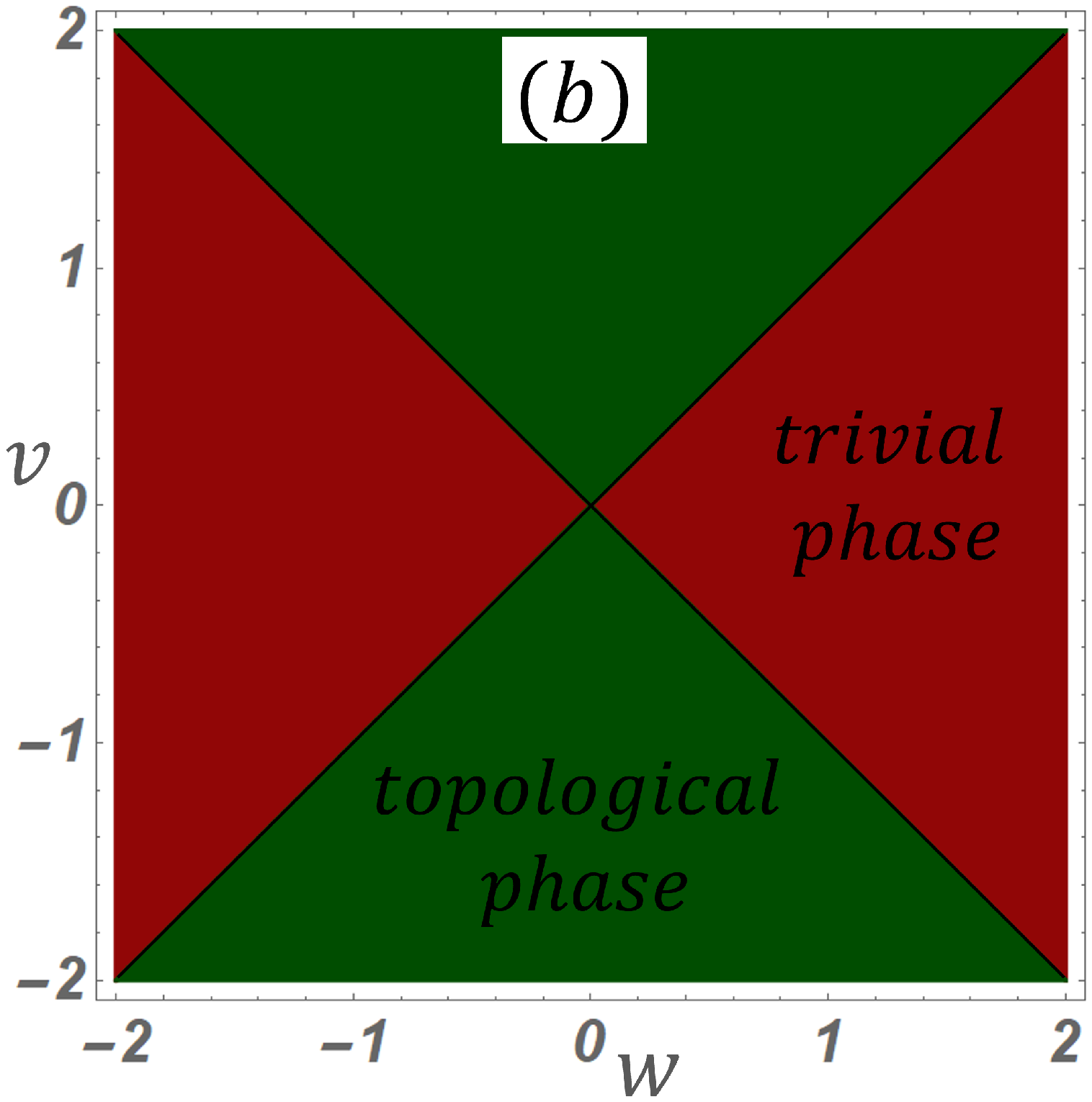}\\
	\vspace{0.5cm}
	\includegraphics[scale=0.25]{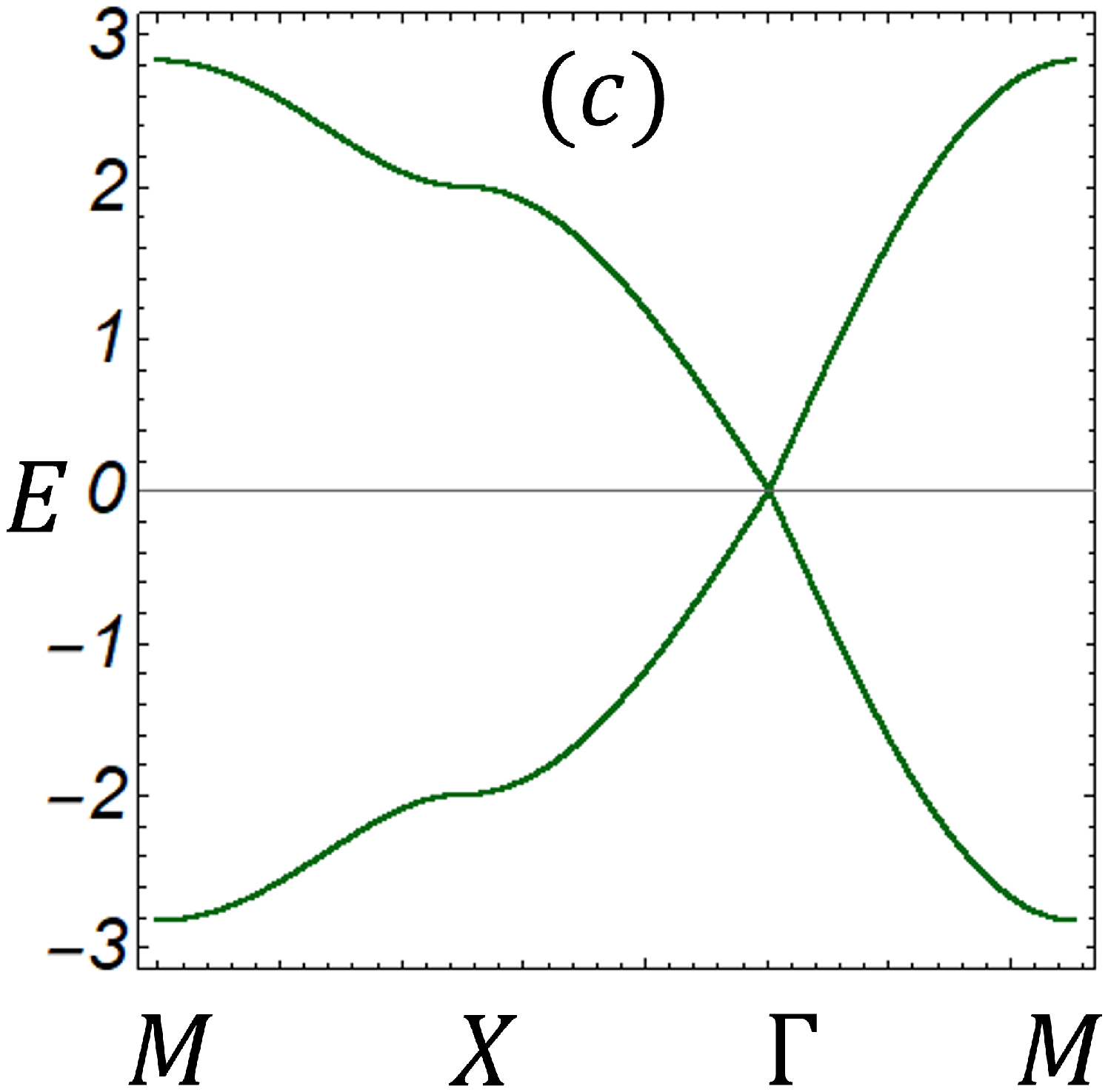}
	\includegraphics[scale=0.25]{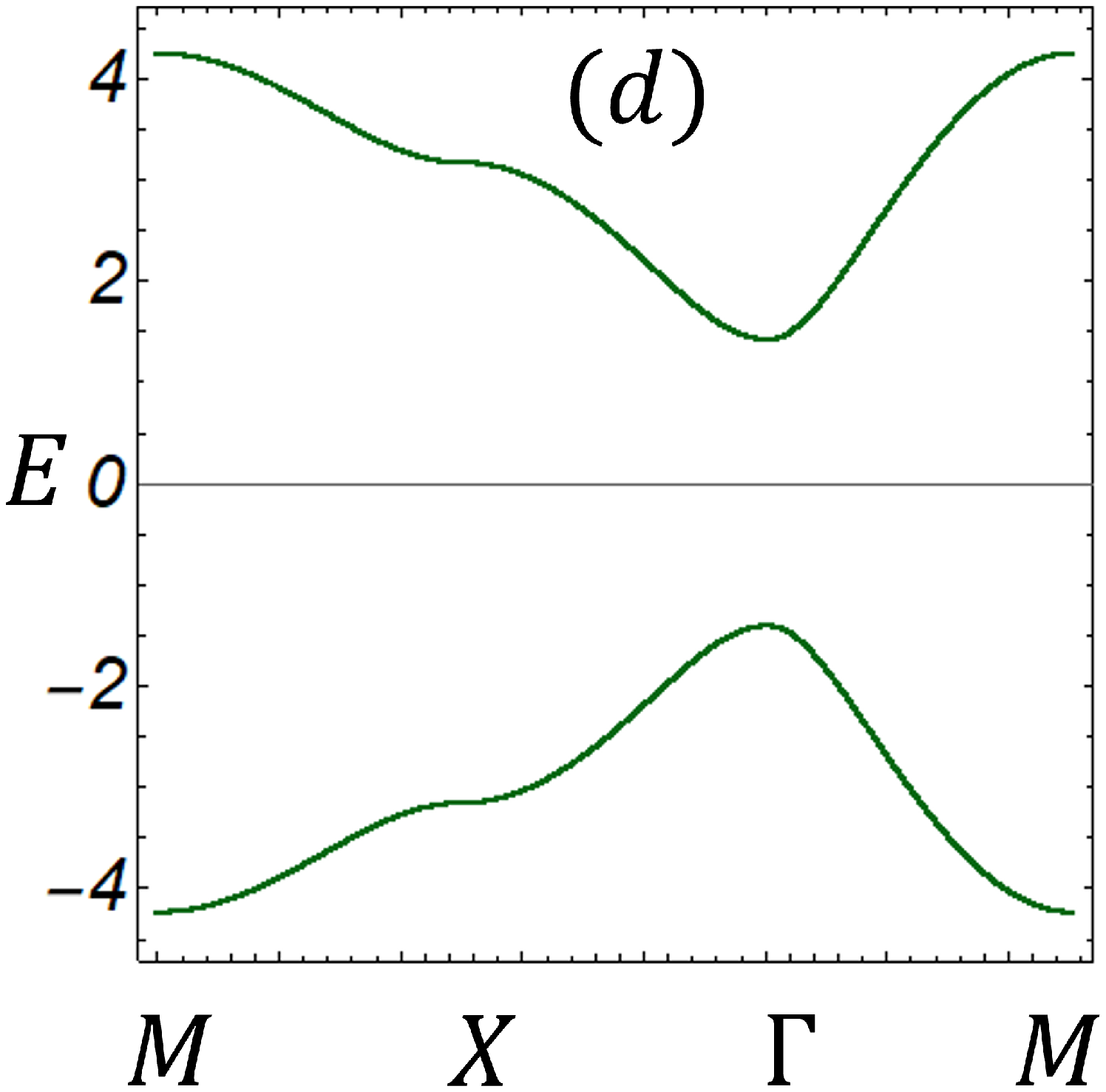}
	\includegraphics[scale=0.4]{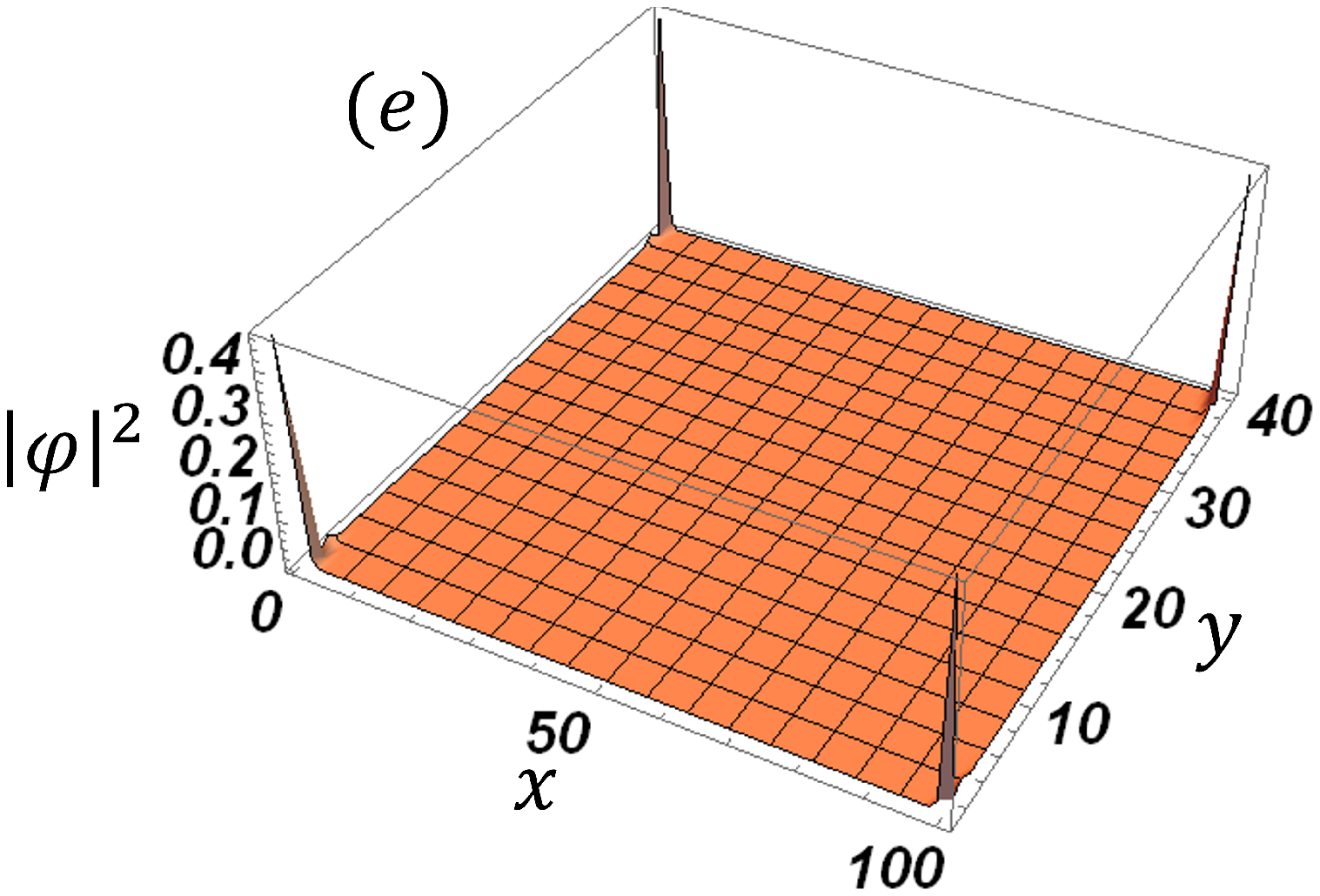}
	\caption{(\textbf{a}) Lattice geometry in the Majorana basis. Blue and grey circles denote, respectively, $a$ and $b$ species of Majorana fermions associated  with a complex fermionic mode. The square plaquette identifies the translational invariant unit cell while $w$ and $v$ are, respectively, the intracell and intercell couplings. A dashed line indicates a negative coupling. (\textbf{b}) Topological phase diagram of the BBH model in the parameter space v, w. Green and red regions identify the topological and trivial phases, respectively. Panel (\textbf{c}) shows the spectrum along the $MX\Gamma M$ path in the Brillouin zone at the phase transition point $w = v = 1$ and panel (\textbf{d}) shows the spectrum in a trivial (topological) phase $w = 2$ , $v = 1 (w = 1, v = 2)$. In panel (\textbf{e}), we show the square modulus of the lowest four energy modes corresponding to a gap closing point ($w=0.2, v=1$).}
	\label{SSHMajoranaModel}
\end{figure*}
A scheme of the lattice is reported in panel (a) of Figure \ref{SSHMajoranaModel}, where the translational invariant unit cell is indicated by the blue square plaquette surrounding the Majorana operators $a$ and $b$ of Equation \eqref{MajoranaH} which, on the other hand, are represented by two circles of different colors. When Mfs operators are expressed in terms of fermion operators, the model reduces to Kitaev chains coupled by a staggered pairing coupling:
\begin{eqnarray}
	\begin{aligned}
		H_0'=&\sum_{m,l=1}^{L,N}\mu c^\dagger_{m,l}c_{m,l}+\sum_{m,l=1}^{L-1,N}(t c^\dagger_{m,l}c_{m+1,l}+\Delta c_{m,l} c_{m+1,l}+h.c.)\\
		&+\sum_{m,l=1}^{L,N-1}(\Delta_1 c_{m,l} c_{m,l+1}+h.c),
	\end{aligned}
	\label{SSHKitaev}
\end{eqnarray}
with $\mu=w$, $t=\Delta=-v/2$ and $\Delta_1$:
\begin{eqnarray}
	\Delta_1=
	\begin{cases}
		-iw,\ \ l=odd\\
		-iv,\ \ l=even.\\
	\end{cases}
	\label{MainstaggeredPair}
\end{eqnarray}
The system belongs to the second-order topological superconductors (HOTSC2) class, featuring corner Majorana states~\cite{condmat6020015}.
The simultaneous presence of crystalline symmetries and standard symmetries ($\mathcal{C}$, $\mathcal{P}$, $\mathcal{T}$) is essential to ensure the realization of second--order topological superconductivity. Indeed, the system satisfies chiral ($\mathcal{C}$), particle-hole ( $\mathcal{P}$) and time reversal ($\mathcal{T}$) symmetries, belonging to the trivial two-dimensional $BDI$ class of the ten-fold classification~\cite{PhysRevB.55.1142}. Nonetheless, robust corner Majorana modes still appear because of the $C_4$ symmetry, i.e. a rotation of $\theta=\pi/2$ exchanging $x \rightarrow y$, $y \rightarrow -x$ and reflection symmetries $m_x$, $m_y$, corresponding respectively to $x \rightarrow -x$, $y \rightarrow -y$.\\
The phase diagram can be obtained by projection of the 2D Zak phase \cite{RevModPhys.66.899,PhysRevLett.118.076803} which is quantized as $\textbf{P}=(1/2,1/2)$ in the topological phase and vanishes in the trivial phase ($\textbf{P}=(0,0)$). On the other hand, the Berry connection vanishes. The phase diagram is reported in panel (b) of Fig. \ref{SSHMajoranaModel}, the green and red colors correspond respectively to the topological and the trivial regime. The band structure along the crystallographic path $M \Gamma X M$, in the first Brillouin zone, features topological phase transition points. Indeed, the spectrum closes when $|w|=|v|$, while it remains gapped both both in the trivial and nontrivial phases when $|w| \neq |v|$ (see panels (c) and (d) of Fig. \ref{SSHMajoranaModel}). In the topological regime the zero-energy Majorana modes are localised at the four corners of the strip geometry and decay exponentially inside the bulk, as shown in panel (e) of Fig. \ref{SSHMajoranaModel}.
\section{Strip of four interacting chains}
\label{InteractingStrip}
In order to go beyond the mean field treatment, we add repulsive Coulomb interactions to the MBBH model with the aim of gaining insight on how the interactions affect the stability of the topological phases, still an open problem despite intense investigations~\cite{10.1088/2053-1583/ac4060,PhysRevB.97.205133,Nat3060,PhysRevB.96.195160}.\\
Here we restrict our analysis to a quasi one-dimensional limit consisting of $N=4$ chains of length $L$. From a computational point of view, the DMRG method~\cite{doi:10.1146/annurev-conmatphys-020911-125018,McCulloch_2007} works at its best when applied to one--dimensional many--body systems associated to gapped Hamiltonians with short--range interactions. For this reason, the simplified limit of four interacting chains analyzed by DMRG techniques based on a tensor--network approach~\cite{ORUS2014117,10.21468/SciPostPhysLectNotes.8,SCHOLLWOCK201196} reduces the computational efforts and increases the power of the numerical algorithms. The strip limit is expected to shed light on the topological properties of the fully two-dimensional Majorana BBH model, which realizes the thermodynamic limit of our strip ($N=2$, $L$).

\subsection{Analytical results}
\label{AnalyticalResults}
We include a local repulsive interaction of strength $U>0$ to the Hamiltonian in Equation (\ref{SSHKitaev}), $H=H_0'+H_I$, where:
\begin{eqnarray}
	\begin{aligned}
		H'_0=&\sum_{m,l=1}^{L,4}\mu c^\dagger_{m,l}c_{m,l}+\sum_{m,l=1}^{L-1,4}(t c^\dagger_{m,l}c_{m+1,l}+\Delta c_{m,l} c_{m+1,l}+h.c.)+\sum_{m,l=1}^{L,3}(\Delta_1 c_{m,l} c_{m,l+1}+h.c)\\
		H_I=&\sum_{m,l=1}^{L-1,4} U(n_{m,l}n_{m+1,l}),
	\end{aligned}
	\label{4ChainsAppendix}
\end{eqnarray}
\begin{figure}
	\centering
	\includegraphics[scale=0.5]{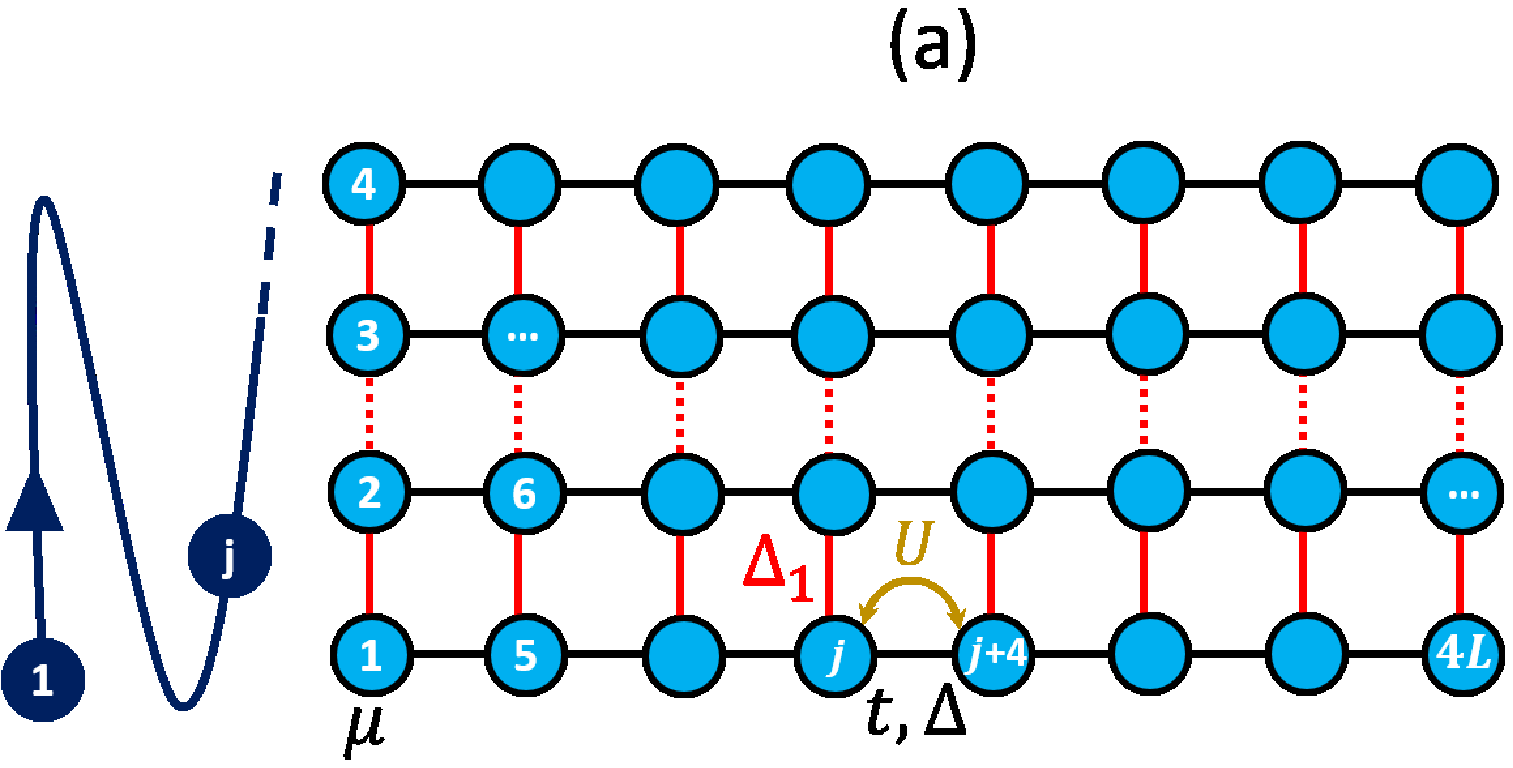}
	\hspace{0.5cm}
	\includegraphics[scale=0.42]{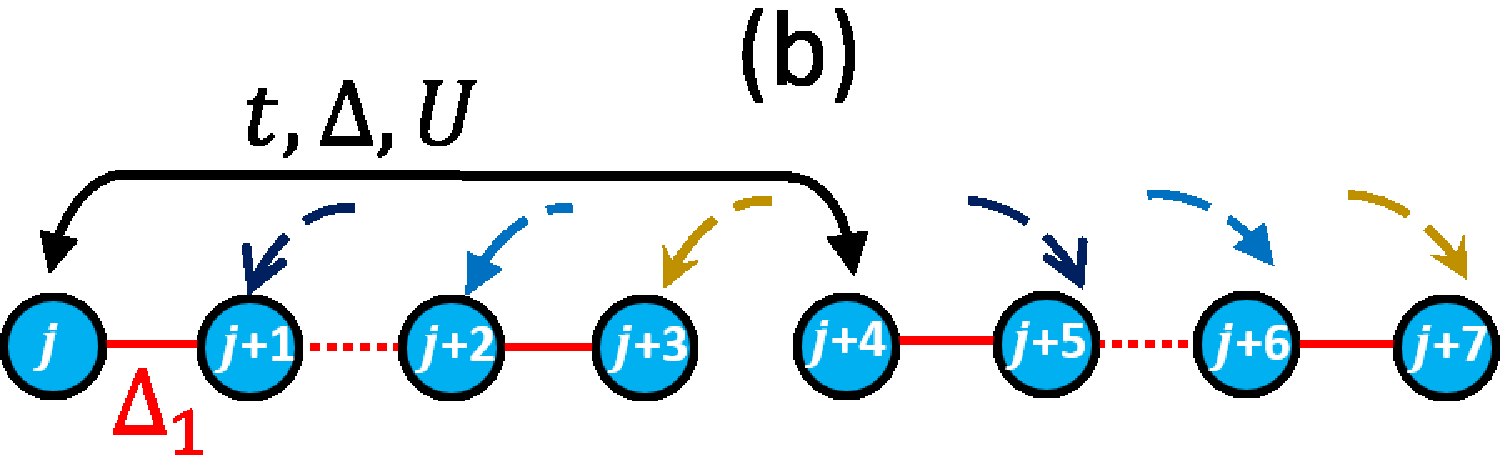}
	\caption{(a) Strip of four chains: the black and red links represent respectively the intrachain and interchain couplings, while the double yellow arrow schematizes the interactions. The numbers inside the circles are the new ordering of the fermionic sites introduced by the mapping to the "curvilinear abscissa" (schematized by the dark blue curve line). (b) One-dimensional model obtained by the mapping. The interchain coupling is promoted to a coupling between nearest neighbouring sites, while the intrachain couplings couple fourth-neighboring sites. In both panels the red links representing $\Delta_1$ correspond to an alternate continuous and dashed line in order to take into account its staggered value (see Equation \ref{MainstaggeredPair}).}
	\label{4chainsStripAppendix}
\end{figure}
with $n_{m,l}=c^\dagger_{m,l} c_{m,l}$. The efficiency of the DMRG algorithn with MPSs for one-dimensional systems~\cite{10.21468/SciPostPhysLectNotes.8} stems from the internal structure of the MPSs state \cite{SCHOLLWOCK201196}. Therefore, we transform the strip geometry of four chains with $L$ sites into a single chain made of $4L$ sites, as shown pictorially in Fig.~\ref{4chainsStripAppendix} panel (a). Introducing a single fermionic species $c_{j}$ ($c_j^\dagger$) and defining a "curvilinear abscissa" which maps the inter-chain interaction terms into nearest-neighbour interaction terms and the intra-chain interaction terms into fourth-neighbour interaction terms, see Fig.~\ref {4chainsStripAppendix} panel (b), the Hamiltonian in Equation (\ref{4ChainsAppendix}) reads:
\begin{eqnarray}
	\begin{aligned}
		H=&\sum_{j=1}^{4L}\mu c^\dagger_{j}c_{j}+\sum_{j=1}^{4L-4}(t c^\dagger_{j}c_{j+4}+\Delta c_{j} c_{j+4}+h.c.)+\\
		&\sum_{j=1}^{4L-1}(\Delta_1 c_{j} c_{j+1}+h.c)+\sum_{j=1}^{4L-4} U(n_{j}n_{j+4}).
	\end{aligned}
	\label{CurvHamiltonianAppendix}
\end{eqnarray}
\begin{figure}
	\centering
	\includegraphics[scale=0.4]{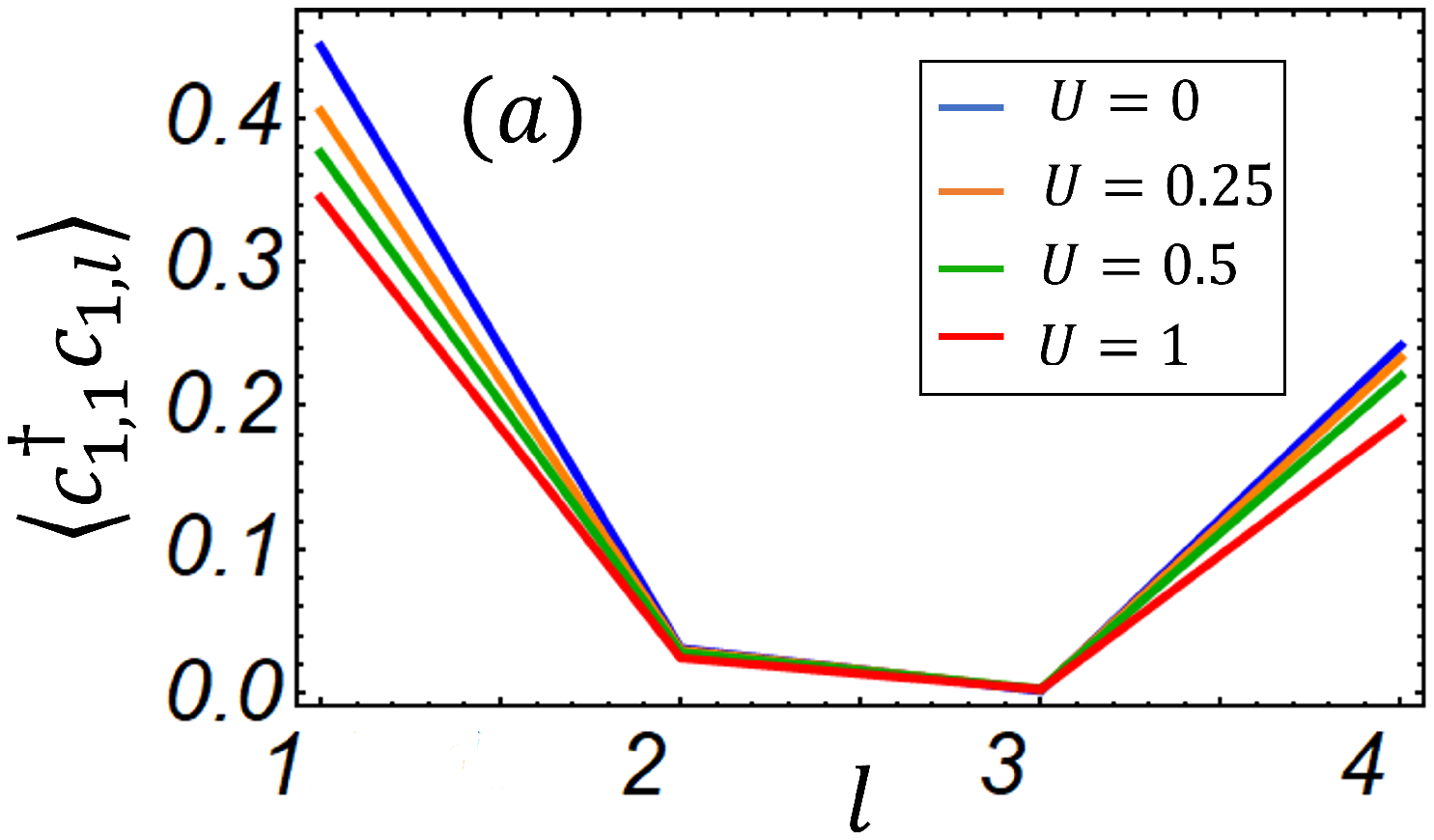}
	\hspace{0.2cm}
	\includegraphics[scale=0.4]{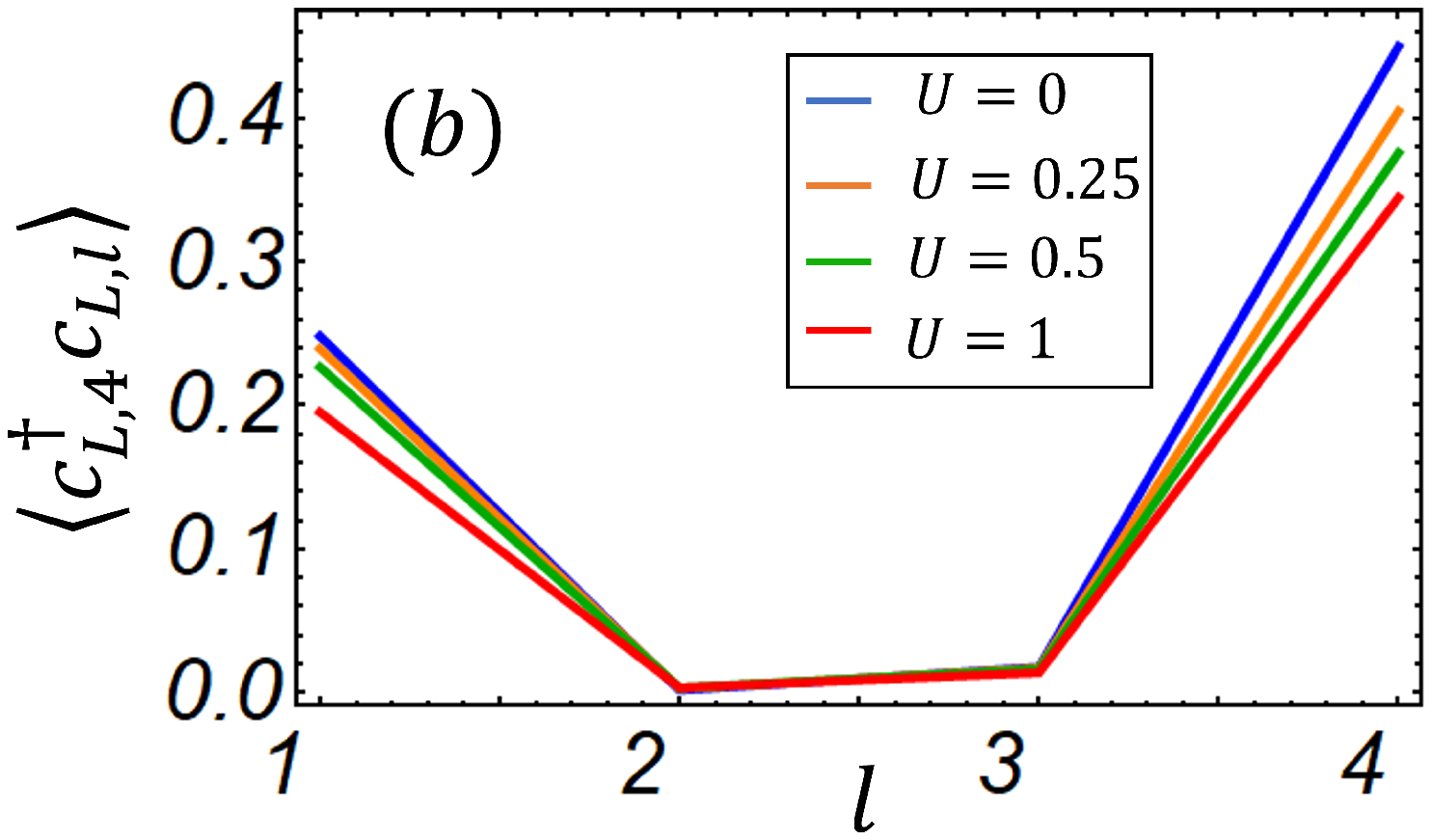}
	\hspace{0.2cm}
	\includegraphics[scale=0.4]{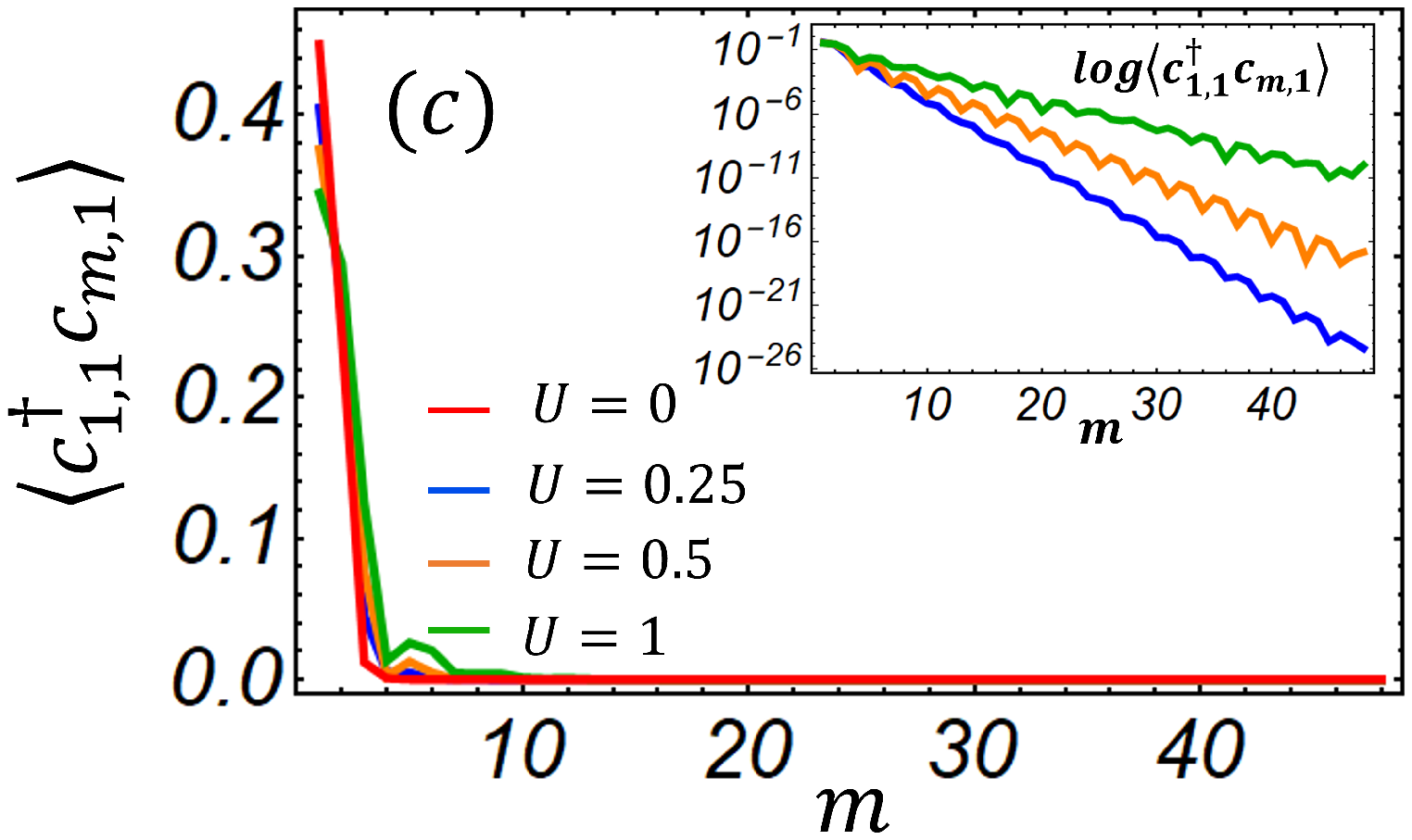}
	\hspace{0.2cm}
	\includegraphics[scale=0.4]{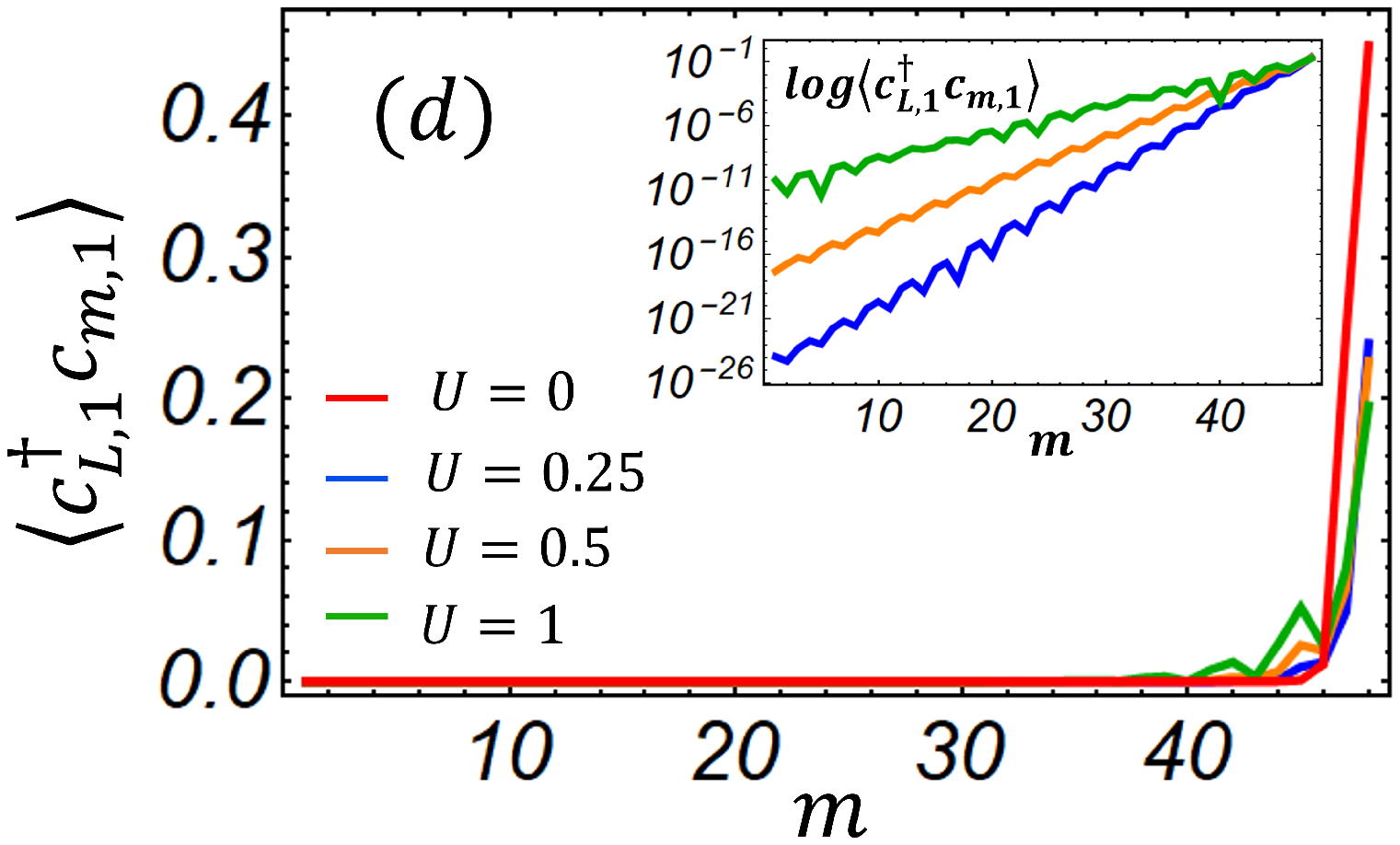}
	\hspace{0.2cm}
	\includegraphics[scale=0.4]{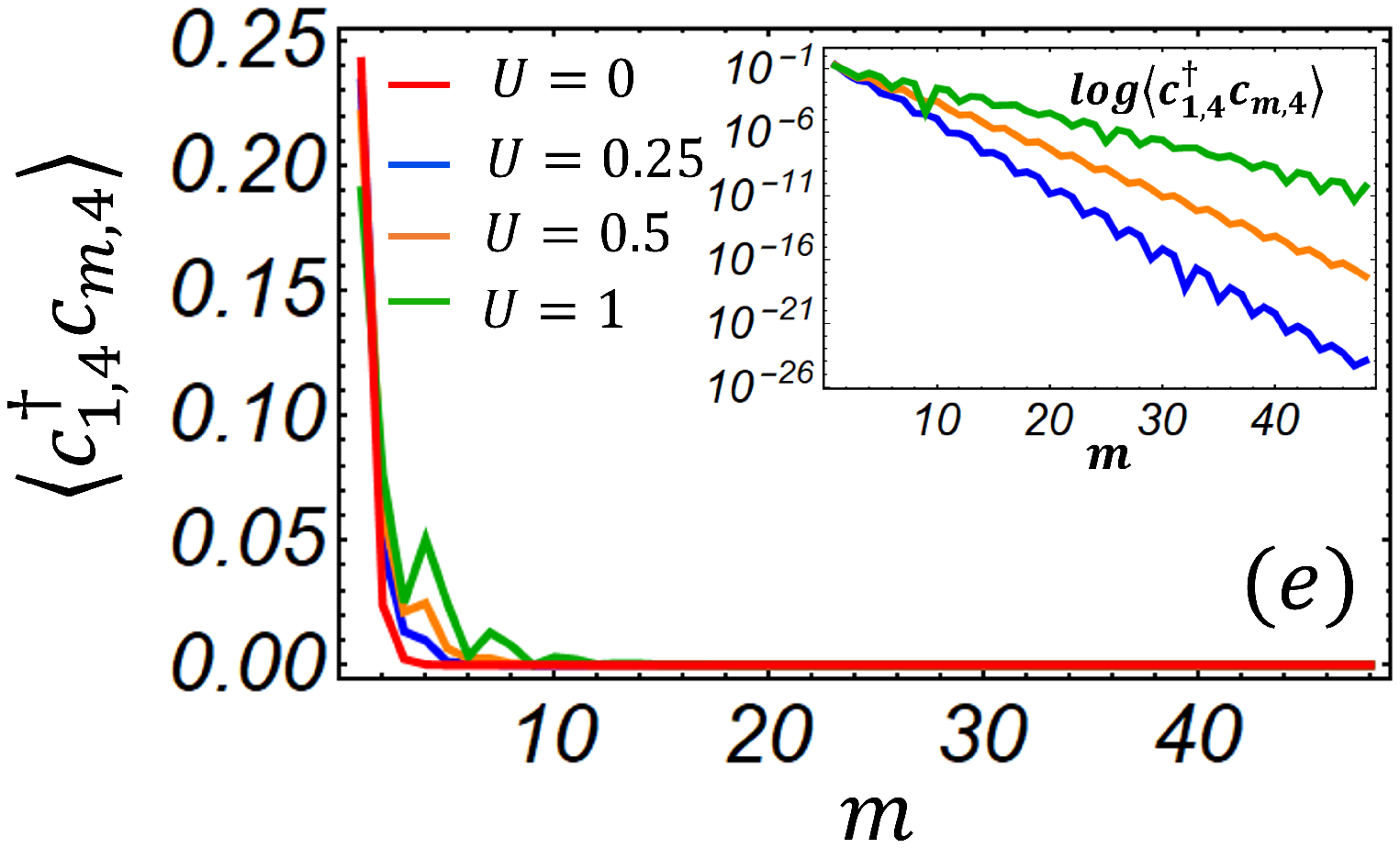}
	\hspace{0.2cm}
	\includegraphics[scale=0.4]{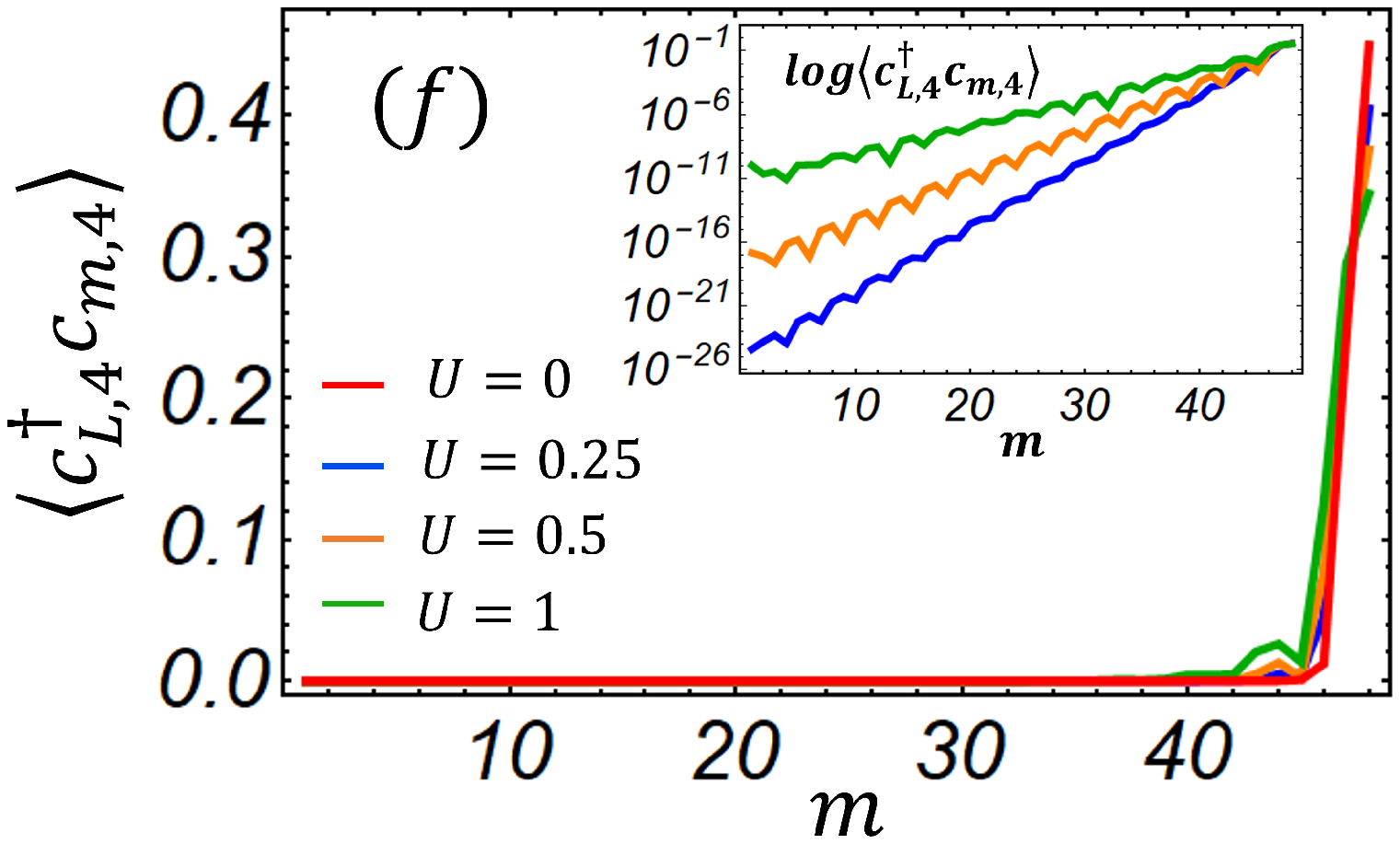}
	\caption{Fermionic correlations as a function of the interaction strength $U$ (in unit of the hopping $v$) $\bra{c_{1,1}^\dagger}\ket{c_{1,l}}$ panel (a), $\bra{c_{L,4}^\dagger}\ket{c_{L,l}}$ panel (b), $\bra{c_{1,1}^\dagger}\ket{c_{m,1}}$ panel (c), $\bra{c_{L,1}^\dagger}\ket{c_{m,1}}$ panel (d), $\bra{c_{1,4}^\dagger}\ket{c_{m,4}}$ panel (e) and $\bra{c_{L,4}^\dagger}\ket{c_{m,4}}$ panel (f), where $l=1,2,3,4$ is the chain index and $m=1$, $\dots$, $48$ the site index. The insets show the exponential penetration of fermionic correlations in logaritmic scale. The model parameters have been fixed as $L=48$, $N=4$, $w=0.1$ and $v=1$. The bond dimension of the DMRG procedure has been fixed as $m=200$.}
	\label{DMRG1}
\end{figure}
\begin{figure}
	\centering
	\includegraphics[scale=0.4]{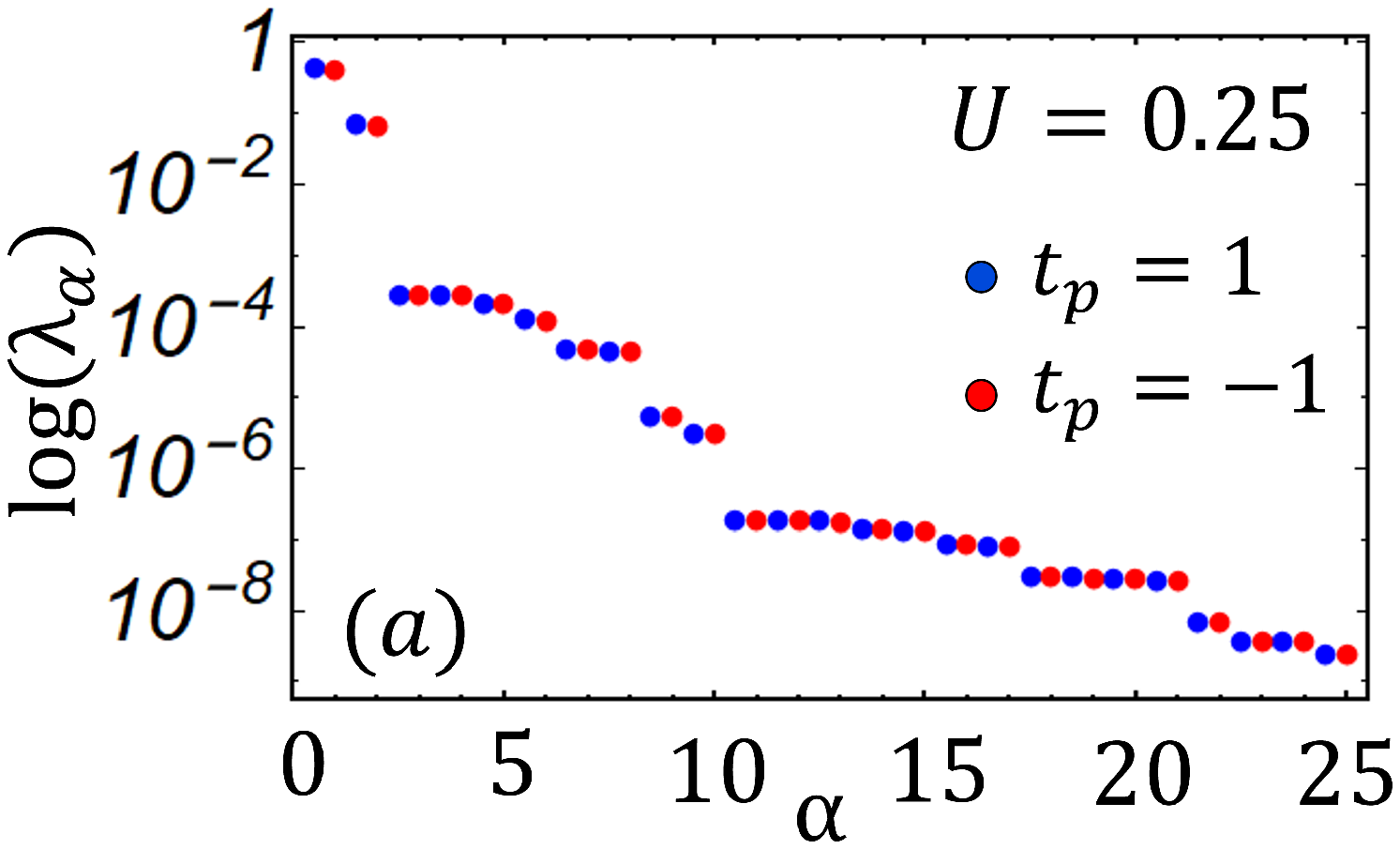}
	\hspace{0.2cm}
	\includegraphics[scale=0.4]{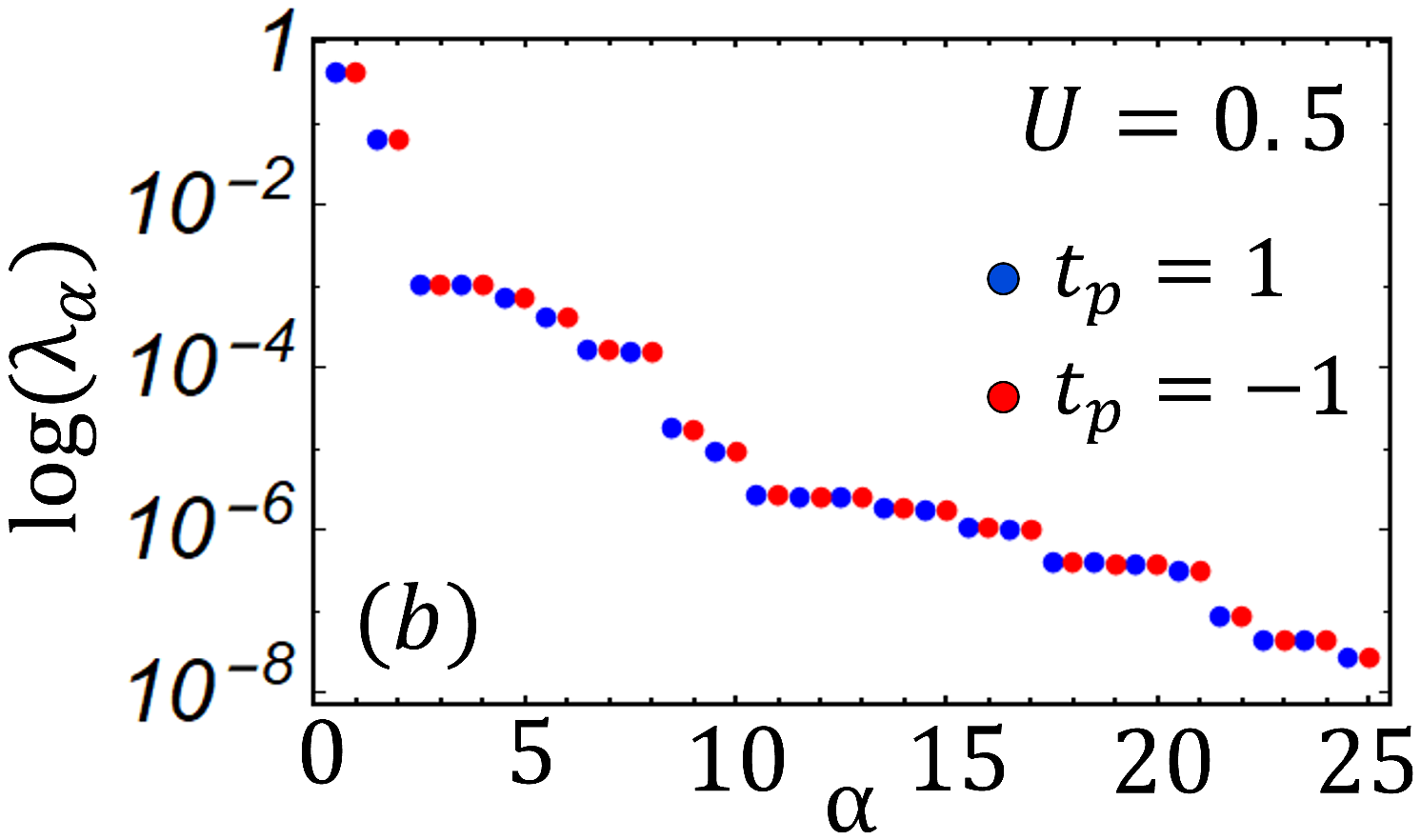}\\
	\vspace{0.3cm}
	\includegraphics[scale=0.4]{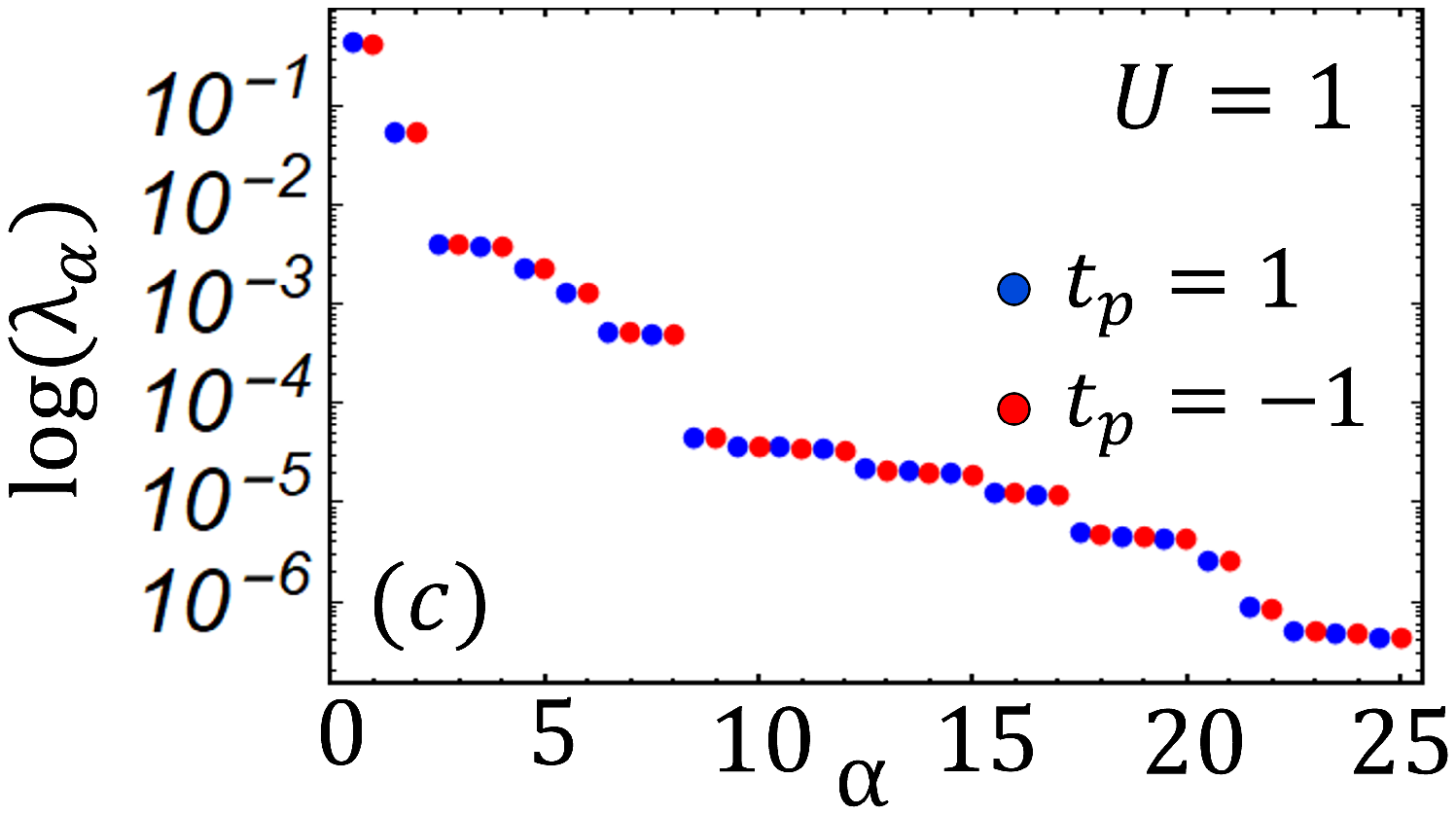}
	\caption{Entanglement spectrum for three different values of the interaction strength: $U=0.25$ panel (a), $U=0.5$ panel (b) and $U=1$ panel (c). The two different parity sectors have been highlighted by the red and blue colours. The model parameters have been fixed as $L=48$, $N=4$, $w=0.1$ and $v=1$. The bond dimension of the DMRG procedure has been fixed as $m=200$. $\alpha$ is the eigenvalue index ($\lambda_\alpha$).}
	\label{DMRG2}
\end{figure}
As fermionic operators of different sites anti-commute, lattice fermionic fields are nonlocal, and it is therefore impossible for such fields to determine a local matrix representation, i.e. a representation in which matrices representing the fermions commute when belonging to different lattice sites. The correct on-site mapping is provided by the Jordan-Wigner transformation, an highly non-local mapping between fermionic operators and spin $1/2$ operators that is a particular case of the general Klein transformation in quantum field theory~\cite{2017}. On each site, an empty fermionic occupation is mapped into an up spin and an occupied one into a down spin. The nonlocal part of this mapping is the so-called Jordan-Wigner string and fixes the (anti)commutation relation between sites, by
counting the parity of overturned sites to the left of the spin on which it is applied. This transformation explicitly breaks the translational invariance of the model, by singling out a particular site as the origin for each lattice string:
\begin{eqnarray}
	\begin{cases}
		c_j=e^{-i \pi \sum_{l=1}^{j-1} c^\dagger_l c_l}\sigma^{+}_j\\
		c^\dagger_j=\sigma^{-}_je^{i \pi \sum_{l=1}^{j-1} c^\dagger_l c_l}\\
		n_j=\frac{1-\sigma^{z}_j}{2},
	\end{cases}
	\label{JWtransformation}
\end{eqnarray}
where the aforementioned string parity of the overturned sites is $e^{-i \pi \sum_{l=1}^{j-1} c^\dagger_l c_l}$. The operators $\sigma_j^{(+,-)}=(\sigma_j^{x}\pm i\sigma_j^{y})/2$ are the well-known combination of Pauli matrices and the last relation in Eq.~(\ref{JWtransformation}) allows to express the parity operator of the fermionic site $j$ as $e^{-i \pi c^\dagger_j c_j}=\sigma_j^{z}$. Using the algebra of the spin-$1/2$ operators and imposing commutation of the Pauli matrices defined on different lattice sites, it is straightforward to show that
\begin{eqnarray}
	\begin{aligned}
		&c_j c_{j+1}=-\sigma^{+}_j \sigma^{+}_{j+1}\\
		&c_j^\dagger c_{j+4}=\sigma^{-}_j \biggr(\prod_{l=j+1}^{j+3} \sigma_{l}^z  \biggl) \sigma^{+}_{j+4}\\
		&c_j c_{j+4}=-\sigma^{+}_j \biggr(\prod_{l=j+1}^{j+3} \sigma_{l}^z  \biggl) \sigma^{+}_{j+4} \, .\\
	\end{aligned}
\end{eqnarray}
The Hamiltonian Eq.~(\ref{CurvHamiltonianAppendix}) in terms of spin-$1/2$ operators becomes:
\begin{eqnarray}
	\begin{aligned}
		H=&\sum_{j=1}^{4L}\mu n_j+\sum_{j=1}^{4L-4}(t \sigma^{-}_j\sigma^{z}_{j+1} \sigma^{z}_{j+2} \sigma^{z}_{j+3}\sigma^{+}_{j+4}-\Delta \sigma^{+}_j\sigma^{z}_{j+1} \sigma^{z}_{j+2} \sigma^{z}_{j+3}\sigma^{+}_{j+4}+h.c.)+\\
		&\sum_{j=1}^{4L-1}(\Delta_1 \sigma^+_j \sigma^+_{j+1}+h.c)+\sum_{j=1}^{4L-4} U n_{j}n_{j+4} \, .
	\end{aligned}
	\label{JWH}
\end{eqnarray}
We see that in Eq.~(\ref{JWH}) there appear explicitly the parity strings between sites that are not nearest neighbours; these strings encode the non-local character of the mapping.\\
Eq.~(\ref{JWH}) can be represented as a matrix product operator (MPO), i.e. as a site by site decomposition of the Hamiltonian into the product of matrices containing operators that act only on one site:
\begin{eqnarray}
	H= \sum_{\{w\},\{\sigma\},\{\sigma'\}} M^{\sigma'_1,\sigma_1}_{w_1} M^{\sigma'_1,\sigma_1}_{w_1,w_2} \dots M^{\sigma'_L,\sigma_L}_{w_{L-1}}\
	\ket{\sigma^{'}_1 \sigma^{'}_2 \dots \sigma^{'}_L} \bra{\sigma^{}_L \sigma^{}_{L-1} \dots \sigma^{}_1},
	\label{MPODef}
\end{eqnarray}
with $M^{\sigma^{'}_k,\sigma_k}_{w_{k-1},w_k}$ a fourth rank tensor.
The physical indices ($\sigma^{'}_k,\sigma_k=1,2$) depend on the dimension of the single-site Hilbert spaces, while the tensor indices ($w_{k-1},w_k=1, \dots, 14$) depend on the structure of Hamiltonian in Eq.~(\ref{JWH}).\\
Not all sites have square MPOs. Indeed, the first and last tensors ($M^{\sigma^{'}_1,\sigma_1}_{w_{1}}$, $M^{\sigma^{'}_L,\sigma_L}_{w_{L-1}}$) are of rank three and are crucial to reproduce the full Hamiltonian, Eq.~(\ref{JWH}).
The analytic expressions of local tensors, the corresponding graph representations of the tensor structure and other technical details are reported and discussed in Appendix \ref{AppendixA}. \\
Once the MPO structure of the Hamiltonian is obtained, simulations with the DMRG code are performed~\footnote{The authors thank Matteo Rizzi for providing them with the DMRG source code.}. The results of the numerical simulations and the emerging physical properties are discussed in the following subsection.

\subsection{Numerical results}
\label{DMRGInteracting}

Topological phases are usuallyt identified according to the following criteria: (i) evidence of degenerate ground states with different parities, (ii) evidence of nonlocal fermionic correlations between the edges, (iii) a degenerate entanglement spectrum~\cite{PhysRevLett.111.173004,PhysRevB.84.014503}. Here we focus on properties (ii) and (iii). We consider a strip of length $L=48$, width $N=4$, and set the parameters of the model at the point of exact topological degeneracy, $w=0.1$, $v=1$, in the phase diagram reported in Fig.~\ref{SSHMajoranaModel} panel (b).\\
In Fig.~\ref{DMRG1} we report the non-local fermion correlations as computed on the many--body ground state $\ket{GS}$. We use the shorthand notation $<c^\dagger_{m,l} c_{m',l'}>$ in place of $\bra{GS}c^\dagger_{m,l} c_{m',l'} \ket{GS}$. In Panels (a) and (b) of Fig.~\ref{DMRG1} we report, respectively,  $<c_{1,1}^\dagger c_{1,l}>$ and $<c_{L,4}^\dagger c_{L,l}>$ which are the fermionic correlations along the short dimension of the strip, while panels from (c) to (f) show, respectively, the correlations of the four corners along the length $L$ of the strip: $<c_{1,1}^\dagger c_{m,1}>$, $<c_{L,1}^\dagger c_{m,1}>$, $<c_{1,4}^\dagger c_{m,4}>$, and $<c_{L,4}^\dagger c_{m,4}>$. Here $l=1,2,3,4$ is the chain index while $m=1$, $\dots$, $L$ denotes the site index. We also draw a comparison between three cases corresponding to three different values of the interaction strength ($U=0.25$, $0.5$, $1$) and the non-interacting case $U=0$.\\
The presence of robust fermionic modes localized at the edge of the strip is clearly visible in Fig.~\ref{DMRG1}. In particular, in panel (a) the correlations attain finite values at corners $(1,1)-(1,4)$ and in panel (b), at corners $(L,1)-(L,4)$, for all the three reported values of the interaction strength. The correlations vanish elsewhere in the strip geometry. Moreover, panels from (c) to (f) show that the fermionic correlations rapidly decay to zero when one of the four corners of the strip is fixed and the site index $m$ is increased from $1$ to $L$. The aforementioned analysis suggests that, for increasing width of the strip $N\gg 4$, these fermionic modes are expected to split into corner Majorana modes, correlating at the corners  $(1,1)-(1,N)$ and $(L,1)-(L,N)$.\\
Let us now discuss the behavior of the entanglement spectrum.
Denote by $\ket{\Psi}=\sum_{\alpha=1}^{D} \lambda_\alpha\ket{\phi_\alpha}_A \ket{\phi_\alpha}_B$ the many--body ground state with respect to some bipartition of the system with $\lambda_\alpha$ real numbers and $D=d^{4L}$ the Hilbert space dimension. Here $d$ denotes the dimension of the single-site Hilbert space, and the set of real eigenvalues $ \{ \lambda_\alpha \}_{\alpha=1}^D$ forms the entanglement spectrum. In the DMRG procedure, this expression can be replaced by $\ket{\Psi}=\sum_{\alpha=1}^{m} \lambda_\alpha\ket{\phi_\alpha}_A \ket{\phi_\alpha}_B$, with the entanglement spectrum reducing to $\{ \lambda_\alpha \}_{\alpha=1}^m$, and the dimension $D$ replaced by a fixed bond dimension $m$.
In Figures \ref{DMRG1}, \ref{DMRG2} the bond dimension has been fixed at the value $m=200$\\.
In the topological phase, the entanglement spectrum is expected to be twofold degenerate with respect to the parity sector, this feature being a precursor of the zero energy Majorana edge states. Indeed  the presence of Majorana edge states implies the occupation of nonlocal fermionic modes allowing such degeneracy with respect to parity. Panels (a), (b) and (c) of Fig. \ref{DMRG2} report the behavior of the entanglement spectrum as a function of the eigenvalue index $\alpha$ for three different values of the interaction strength $U$. All panles show a degenerate entanglement spectrum, a typical signature of an ordered phase.\\
Summing up, the findings reported in Fig.~\ref{DMRG1} and Fig.~\ref{DMRG2} show that Coulomb repulsive interactions do not affect significantly the topological phases of the model, featuring robust edge fermionic modes and a degenerate entanglement spectrum also in the strongly interacting regime.

\section{Conclusions}
\label{Conclusions}
We have analyzed the effect of nearest-neighbor Coulomb repulsive interactions on a Majorana BBH model restricted to a strip of $N=4$ chains of length $L$ with interactions along the chains. The simulations, performed using a DMRG procedure, provide some interesting insights on the robustness of topological phases against repulsive interactions. In particular, we have observed robust fermionic modes localized at the edges of the strip, together with a degenerate entanglement spectrum. Both obeservations point at the existence of a robust topological order, persistent even in the regime of strong interactions. These results can be considered preliminary to the investigation of the robustness of Majorana corner states in an interacting two--dimensional MBBH model which can be simulated by a tree tensor network approach in the limit of large lattices.\\
In order to discriminate unambiguously the topological order from other types of order, we will need to go beyond the entanglement spectrum, since the latter does not discriminate between different types of order, e.g. topological order and orders associated to spontaneous symmetry breaking. We will thus need to consider entanglement measures able to quantify the nonlocal correlations between the edges and, in particular, the long-distance topological entanglement between corner Majorana modes. In fact, such a measure, the topological squashed entanglement, exists and has been recently introduced and successfully applied~\cite{squashed} to the unambiguous characterization of topological order in some basic models of topological superconductivity, including the Kitaev chain, the two-leg Kitaev ladder~\cite{ladder}, and the Kitaev tie~\cite{tie}. We plan to report in the near future the results of a similar investigation along the same lines for the interacting Majorana BBH model.

\appendix
\section{Matrix Product Operator}
\label{AppendixA}

\begin{figure}
	\centering
	\includegraphics[scale=0.6]{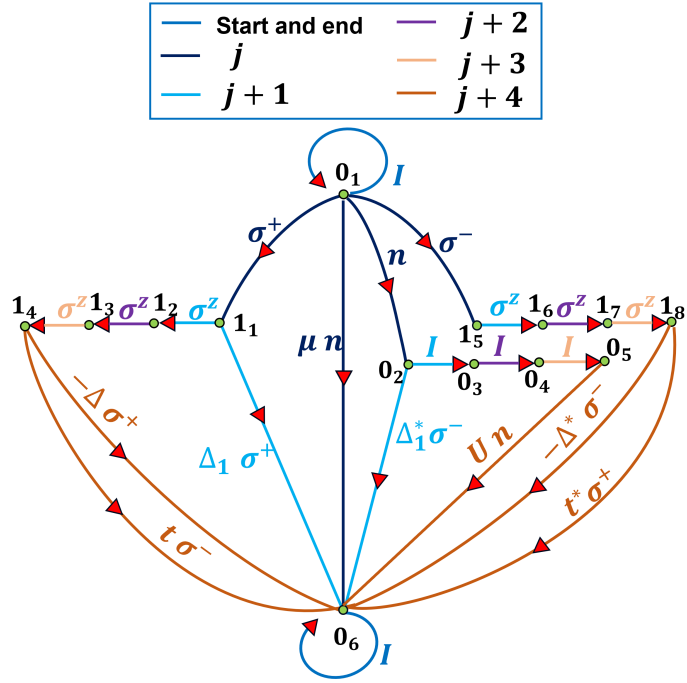}
	\caption{Structure of the graph describing the MPO in 
		Eq.~(\ref{MPOmatrix}). The numbers are the parities of the graph knots, the  subscripts count the degeneracy of such parities and the lines represent the action of the Pauli operators with the properly assigned couplings. A pair of numbers identifies the position of a block matrix element in the MPO, whose value is given by the appropriate Pauli operator, associated with a colored line. Different colours of the lines refer to nearest-, next to nearest-, third-, and fourth-neighbouring interactionterms. The symbol $I$ stands for the identity operator.}
	\label{MPOGRAPH}
\end{figure}

As discussed in the main text, the Hamiltonian of Eq.~(\ref{JWH}) can be seen as a tensor with  $N$ covariant and $N$ contravariant indices and can be factorized into a contracted product of smaller tensors, each carrying one of the original contravariant and covariant indices each, as well as ?bond indices? connecting to the neighbouring factor tensors (See Eq.~(\ref{MPODef})). \\
In some interesting cases, symmetries can also be implemented in the tensors, yielding a twofold benefit: they provide a substantial computational
speed-up, and they allow for precise targeting of symmetry sectors \cite{10.21468/SciPostPhysLectNotes.8}. Our model conserves the parity $\mathcal{Z}_2$ and we thus write the MPOs by targeting the parity sectors. The analytical expressions for all the MPOs when $j=2, \dots L-1$ are:
\begin{eqnarray}
	\mathcal{M}=\begin{blockarray}{ccccccccccccccc}
		&\color{blue}0_1&\color{blue}0_2 &\color{blue}0_3 & \color{blue}0_4&\color{blue}0_5& \color{blue}0_6&\color{red}1_1&\color{red}1_2&\color{red}1_3&\color{red}1_4&\color{red}1_5&\color{red}1_6&\color{red}1_7&\color{red}1_8&\\
		\begin{block}{c(cccccc|cccccccc)}
			\color{blue}0_1&\mathcal{I}&n&0&0&0&\mu n&\sigma^+&0&0&0&\sigma^-&0&0&0\\
			\color{blue}0_2&0&0&\mathcal{I}&0&0&0&0&0&0&0&0&0&0&0\\
			\color{blue}0_3&0&0&0&\mathcal{I}&0&0&0&0&0&0&0&0&0&0\\
			\color{blue}0_4&0&0&0&0&\mathcal{I}&0&0&0&0&0&0&0&0&0\\
			\color{blue}0_5&0&0&0&0&0&Un&0&0&0&0&0&0&0&0\\
			\color{blue}0_6&0&0&0&0&0&\mathcal{I}&0&0&0&0&0&0&0&0\\
			\cline{2-15}
			\color{red}1_1&0&0&0&0&0&\Delta_1 \sigma^+&0&\sigma^z&0&0&0&0&0&0\\
			\color{red}1_2&0&0&0&0&0&0&0&0&\sigma^z&0&0&0&0&0\\
			\color{red}1_3&0&0&0&0&0&0&0&0&0&\sigma^z&0&0&0&0\\
			\color{red}1_4&0&0&0&0&A^+&0&0&0&0&0&0&0&0\\
			\color{red}1_5&0&0&0&0&0&\Delta_1^* \sigma^-&0&0&0&0&0&\sigma^z&0&0\\
			\color{red}1_6&0&0&0&0&0&0&0&0&0&0&0&0&\sigma^z&0\\
			\color{red}1_7&0&0&0&0&0&0&0&0&0&0&0&0&0&\sigma^z\\
			\color{red}1_8&0&0&0&0&0& A^-&0&0&0&0&0&0&0&0\\
		\end{block}
	\end{blockarray}
	\label{MPOmatrix}
\end{eqnarray}
with $A^+=t \sigma^--\Delta \sigma^+$, $A^-=t \sigma^+- \Delta \sigma^-$, $\mathcal{I}$ the identity operator, and where we are grouping the Pauli matrices by parity sector. The result is a block matrix with the superscripts and side-scripts labeling the different parity sectors (blue and red symbols).\\
Every tensor $\mathcal{M}$, due to an excessively complex structure is usually visualized in terms of a graph~\cite{SciPostPhys.3.5.035,Pirvu_2010} describing the expressions of the matrix.
Indeed, in our graph in Fig.~\ref{MPOGRAPH}, the numbers associated with the knots are the parity sectors which can possibly change depending on the action of the specific Pauli operator (corresponding to a colored line).\\
Different colored lines are used to identify nearest-, next to nearest-, third-, and fourth-neighboring interaction terms. The subscripts identify the degeneracy of the corresponding parity sectors, while pairs of numerical indices identify the position of an element of the MPO matrix, whose value is given by the appropriate coupling and Pauli operator.

{\it Acknowledgments} -- We would like to thank Matteo Rizzi and Francesco Romeo for useful discussions. This work was supported by the project Quantox Grant Agreement No. 731473, QuantERA-NET Cofund in Quantum Technologies, implemented within the EU-H2020 Programme, and by MUR (Ministero dell'Universit\'a e della Ricerca) via the project PRIN 2017 "Taming complexity via QUantum Strategies: a Hybrid Integrated Photonic approach" (QUSHIP) Id. 2017SRNBRK.

{\it Author contributions} --Conceptualization, A.M. and R.C.; methodology, R.C.and F.I.; software, A.M.; formal analysis, A.M.; writing---original draft preparation, A.M., R.C. and F.I.; writing---review and editing, R.C. and F.I.; funding acquisition, R.C. and F.I. . All authors have read and agreed to the published version of the manuscript.
\bibliography{Bibtex_new}
\end{document}